\documentclass[aps,prl,twocolumn,superscriptaddress,floatfix,showpacs]{revtex4-1}

\usepackage{graphicx,amsmath,amsfonts,amssymb,color,ulem,bm,tabularx}
\usepackage{float}
\def\be{\begin{equation}}
\def\ee{\end{equation}}
\def\betaDPETsc{$\beta''$-(ET)$_2$SF$_5$CH$_2$CF$_2$SO$_3$}

\def\betaDPao{$\beta''$-(BEDT-TTF)$_2$SF$_5$CH$_2$CF$_2$SO$_3$}
\def\betaDPaoSC{$\beta''$-(BEDT-TTF)$_2$SF$_5$CH$_2$CF$_2$SO$_3$}

\def\betadp{$\beta''$}
\def\betaDP{$\beta''$}

\def\t1{$T_1^{-1}$}

\def\iTone{$T_1^{-1}$}

\def\C{$^{13}$C}
\def\iTone{$T_1^{-1}$}
\def\iToneT{$[T_1T]^{-1}$}

\begin{document}
\title{Charge fluctuations and superconductivity in organic conductors: the case of \betaDPaoSC}

\author{G.~Koutroulakis}
\affiliation{Department of Physics $\&$ Astronomy, UCLA, Los Angeles, CA 90095, USA}

\author{H.~K\"uhne}
\affiliation{Hochfeld-Magnetlabor Dresden (HLD-EMFL), Helmholtz-Zentrum Dresden-Rossendorf, D-01314 Dresden, Germany}

\author{H.-H.~Wang}
\affiliation{Department of Materials Science and Engineering, UCLA, Los Angeles, CA 90095, USA}

\author{J.~A.~Schlueter}
\affiliation{Materials Science Division, Argonne National Laboratory, Argonne, Illinois 60439, USA} \affiliation{Division of Materials Research, National Science Foundation, Arlington, VA 22230 USA}

\author{J.~Wosnitza}
\affiliation{Hochfeld-Magnetlabor Dresden (HLD-EMFL), Helmholtz-Zentrum Dresden-Rossendorf, D-01314 Dresden, Germany}
\affiliation{Institut f\"{u}r Festk\"{o}rperphysik, TU Dresden, D-01069 Dresden, Germany}

\author{S.~E.~Brown}
\affiliation{Department of Physics $\&$ Astronomy, UCLA, Los Angeles, CA 90095, USA}

\voffset=0.5cm

\begin{abstract}
A \C\ NMR study of the normal and superconducting states of the all-organic charge-transfer salt \betaDPao\ is presented. We find that the normal state is a charge-ordered metal configured as vertical stripes, produced by a combination of 1/4-filling, correlations, and a polar counterion sublattice. The NMR properties associated with the superconducting state are consistent with gap nodes and singlet pairing, and therefore similar to other organic superconductors. Quite distinct, however, is the absence of evidence for low-energy antiferromagnetic spin fluctuations for $T>T_c=4.5$ K. Both aspects are discussed in the context of a proposal that the pairing in this compound is driven by charge fluctuations.

%organic superconductors, NMR in superconductors, pairing symmetries, superconducting phase diagrams
\end{abstract}

\maketitle

%introduction
There are currently more than 50 known distinct superconducting charge transfer salts based on the bisethylenedithio-tetrathiafulvalene (BEDT-TTF, or ET) donor \cite{Ishiguro:1998,Lebed:2008}. In all of them, the ET molecules assemble in sheets separated by layers of negatively-charged counterions, most commonly in the ratio of 2:1, so that singly-charged counterions leave behind 1/2 delocalized hole per donor. The combination of relatively low charge-carrier density and quasi-two dimensional (q2D) electronic structures enhance the influence of correlations. Thus, the superconductors are located proximate to correlated insulating states, where tuning between insulating and superconducting states is typically controlled by changing the lattice constants. Among the most familiar examples are the closely-related q1D (TMTSF)$_2X$ (Bechgaard) salts, and the q2D $\kappa$-(ET)$_2X$ compounds \cite{Ardavan:2012,Brown:2015}, where the insulators are also antiferromagnetic (AF) or spin liquid \cite{Kurosaki:2005}. A broader interest in these compounds derives from the proposition, as first articulated by Emery \cite{Emery:1983} and motivated in part by the proximity to AF ground states, that the superconductivity appearing on the high-pressure side of the metal-insulator transition involves magnetically mediated pairing. Taking the issue further is the question of whether the magnetic mechanism not only applies, but is ubiquitous amongst molecular superconductors.

The predominance of the Mott state in the $\kappa$ materials follows from the dimerized intralayer arrangement of the ET molecules \cite{Oshima:1988}, such that the system is effectively 1/2-filled \cite{Kino:1996}. Whereas, in practice if not precisely by symmetry, the $\theta$ and $\beta''$ configurations are considered 1/4-filled. Nevertheless, a strong tendency for charge ordering (CO) \cite{Seo:2006,Takahashi:2006} produces insulating ground states far more commonly than superconducting ground states \cite{Mori:1998,Mori:1998b}, in accordance with the applicability of an extended Hubbard model which includes nearest-neighbor repulsive interactions $V$ as well as on-site $U$ and hopping parameter $t$. 

A natural question to ask is, does a superconducting state (SC) emerge from the collapsed CO insulating state, and to what extent are charge fluctuations (CF) relevant to pairing in that case? A variant of this question has been raised recently in the context of underdoped cuprates, where incommensurate CO leads to Fermi-surface reconstruction \cite{Sebastian:2014}. In relation to the 1/4-filled molecular superconductors, calculations producing superconductivity out of the collapsed CO state are mostly limited to mean-field theory. For example, in Ref. \cite{Merino:2001}, fluctuations remanent of the checkerboard CO phase are shown, in a slave-boson approach, to stabilize %\sout{in mean-field}
a $d_{xy}$ superconducting state on the square lattice. Though limited in that case to $U/t\to\infty$, $d_{xy}$ SC is also produced in the random phase approximation (RPA) for finite $U/t$ proximate to both spin and charge density wave (SDW and CDW) instabilities, also on the square lattice \cite{Kobayashi:2004}, which are stabilized in the large $V$ and $U$ limits, respectively. %\sout{The insulating states are associated with the repulsive interactions $U, V$; the CDW state is dominant in large $V$ and SDW for large $U$.} %\sout{The superconducting phase is contiguous between these limits, around the edge of the boundary to the insulating states.}
Although applicability to the real materials is unclear, the RPA results indicate that to consider pairing by pure CF is probably an oversimplification.

\begin{figure}
\includegraphics[width=3in]{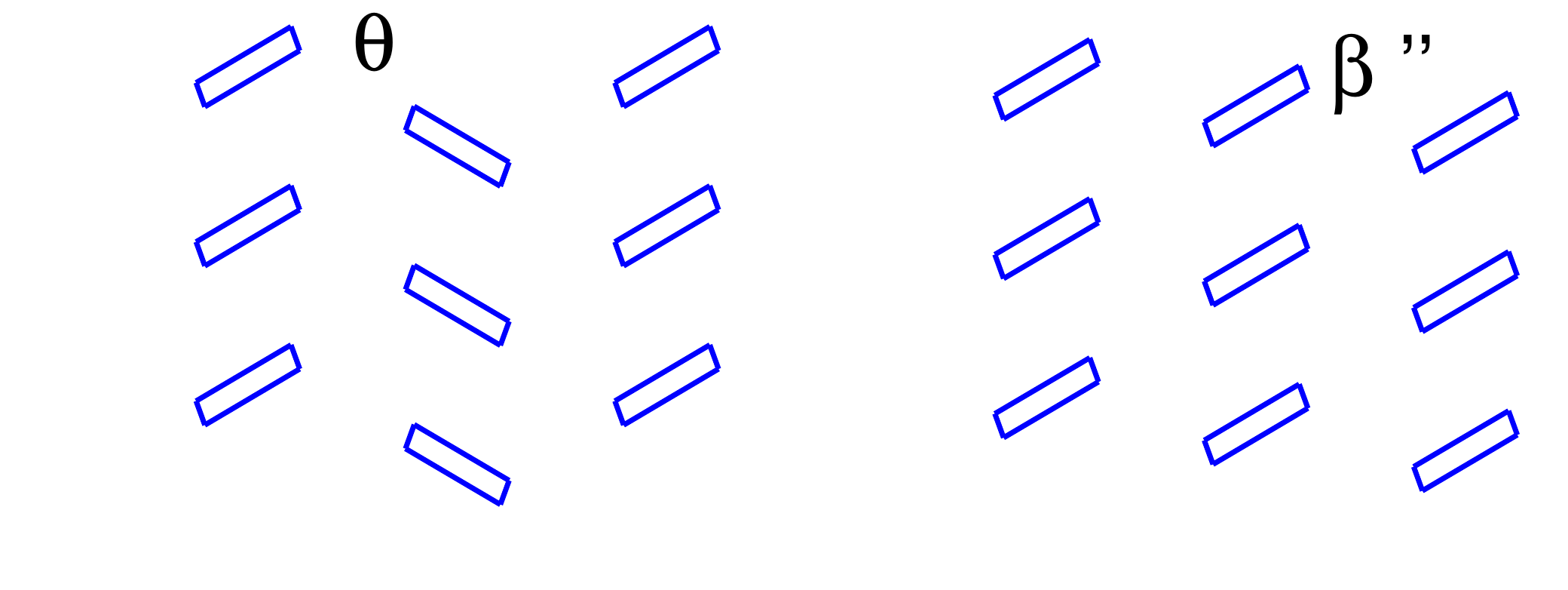}
\caption{Representation of the planar molecular donor arrangements of the $\theta$ and $\beta''$ salts. The long axis of the ET molecules are into the page.}\label{fig:ThetaBetaDP}
\end{figure}

The so-called $\theta$ as well as the $\beta''$ \cite{Girlando:2014} compounds are cited in Ref. \cite{Merino:2001} as good candidates for testing the idea of CF-mediated superconductivity. In both cases, the molecular layers are comprised of side-by-side stacks of ET molecules (see Fig. \ref{fig:ThetaBetaDP}). The unit cell of \betaDP\ is comprised of 4 ET's donating two electrons to the counterion sublattice. However, with near-neighbor repulsive interactions intrastack and interstack, the systems are presumed unstable to insulating CO states. (This is the case for most of the $\theta$ compounds as well, for example $\theta$-(ET)$_2X$, $X=$RbZn(SCN)$_4$, CsZn(SCN)$_4$, though $X$=I$_3$ is superconducting \cite{Mori:1998,Mori:1999}.) For the $\beta''$ compound considered here, different ground states result with small changes to the counterion which we abbreviate as $X$=SF$_5R$SO$_3$ \cite{Schlueter:2001,Girlando:2014}. For example, $R$=CHFCF$_2$ undergoes a metal-insulator transition (\textit{presumed} at about 170 K, and $R$=CHF is an insulator already at ambient temperature. To make more concrete the idea that the collapse of charge order leads to fluctuations and superconductivity, a general phase diagram has been proposed, whereby the insulators could be tuned with application of pressure so as to produce superconductivity \cite{Girlando:2014}. The superconducting $R$=CH$_2$CF$_2$ salt is placed adjacent to the CO phase boundary because, for example, a strongly temperature-dependent oscillator strength in the mid-infrared conductivity \cite{Kaiser:2010} was interpreted as evidence for CF.

\begin{figure}[htb]
\includegraphics[width=3.3in]{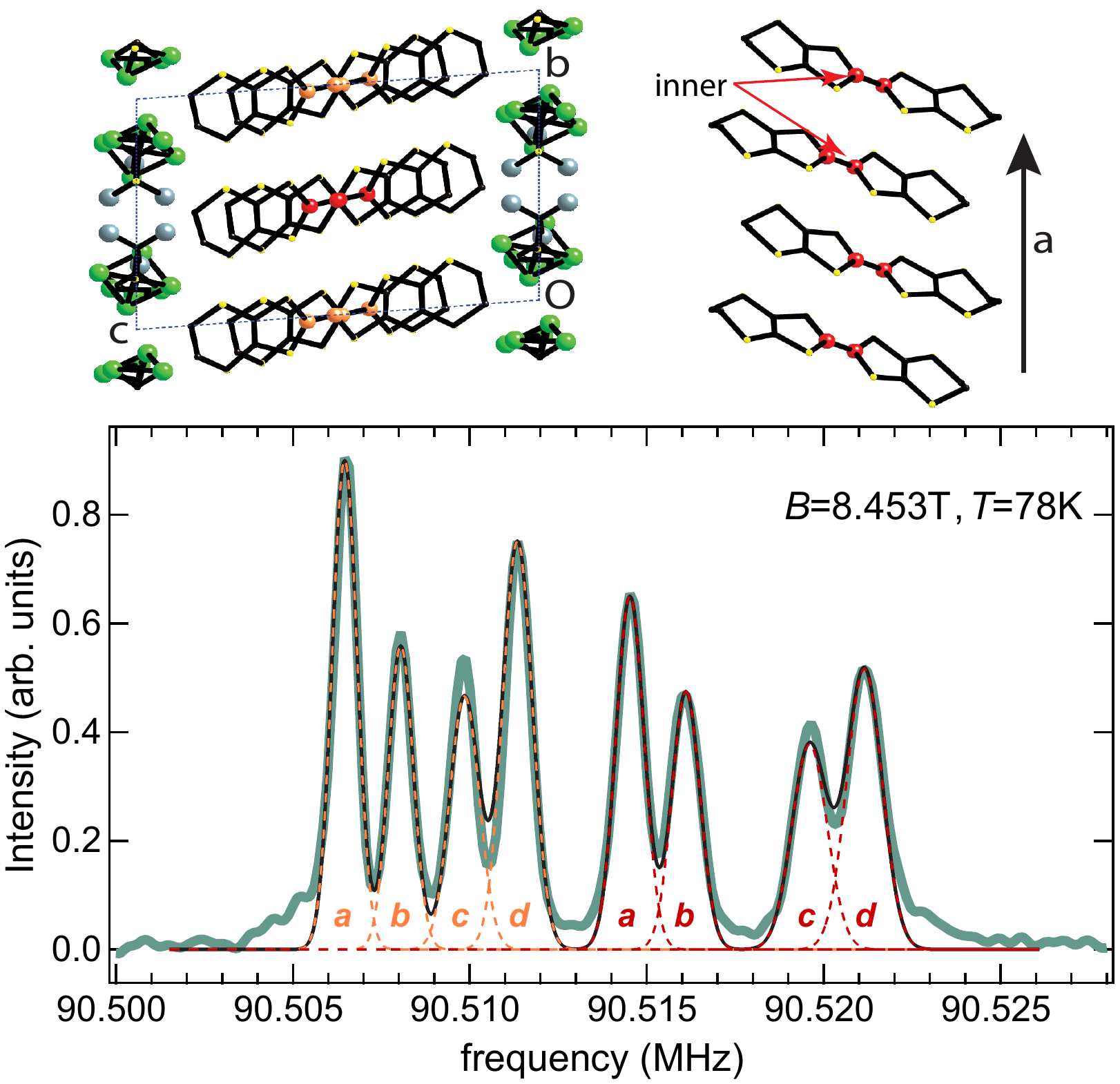}
\caption{\textit{Top:} Crystal structure of \betaDPETsc, viewed along the stack direction $a$ (left), and along the $b$-axis (right). The ET molecules are spin-labelled with \C\ on the bridging sites for the NMR measurements; these are distinguished here by the red and orange colors, which also highlights the crystallographic inequivalence of the two stacks. Within each molecule are inequivalent sites ``inner'' (sites $a,b$) and ``outer'' ($c,d$). From the spin-lattice relaxation (below), we estimate the charge-carrier imbalance between the two stacks is $\rho^{red}/\rho^{orange}\sim1.9$ at $T=78$ K. \textit{Bottom:} \C\ spectrum, which resolves the contributions from red (4 highest-frequency peaks) and orange sites (see text). Dashed lines are simulated peaks per Eq. \ref{eq:SpectralFit}, appropriately broadened, and black solid line is the sum.}\label{fig:betaDPstructure}
\end{figure}

Here, we describe $^{13}$C NMR spectroscopy and relaxation of the $\beta''$ superconductor ($R$=CH$_2$CF$_2$), and compare those results to expectations for %\sout{superconducting}
pairing mediated by CF associated with the ``nearby'' insulating phase of $R$=CHFCF$_2$. The results reveal two independent and inequivalent molecules over the entire temperature range of the measurements, $1.5-300$ K. We associate the two species to the ET molecules in the two structurally independent stacks \cite{Schlueter:2001,Koutroulakis:2015}. Specifically, our sensitivity derives from the hole-density imbalance between the two stacks, which is reminiscent of a vertical-stripe CO configuration \cite{Seo:2006} though the symmetry between stacks is already broken in the crystal (Fig. \ref{fig:betaDPstructure}). The increase of the relative difference of the paramagnetic shifts and relaxation of the two species upon cooling is evidence that the disproportionation is not trivially determined by the polarity of the counterion sublattice, and for that reason we interpret it as a charge-ordered metallic state. The superconductivity emerging from the CO metal is a spin singlet, since it exhibits a diminishing Knight shift for $T<T_c$. This is accompanied by a sharp decrease in the relaxation rate which closely follows the oft-observed \iTone$\sim T^3$, and without any discernible Hebel-Slichter enhancement below $T_c$, even for applied fields $B<$1 T. Most notable, however, is the absence of evidence for antiferromagnetic spin fluctuations (SF) in the normal state, especially when compared to the standard-bearer for spin-fluctuation-mediated organic superconductivity, (TMTSF)$_2X$ \cite{Brown:2008}, the $\kappa$-phase ET superconductors \cite{DeSoto:1995} (which exhibit the highest $T_c$'s amongst ET superconductors), or other correlated superconductors such as cuprate \cite{Takigawa:1991}, heavy-fermion \cite{Zheng:2001}, or iron-based superconductors \cite{Ning:2009}. The expected absence of normal state SF was proposed as a test for CF-mediated superconductivity \cite{Merino:2001}.

The crystals were grown by the standard electrolysis method using ET molecules with \C\ spin-labelled on the central carbons. X-ray diffraction measurements confirmed the known unit-cell parameters of the $\beta''$ phase. Superconductivity was observed at temperatures below $T_c=4.5$ K by dc-susceptibility methods (zero-field cooled) as well as by NMR spin-lattice relaxation and tank-circuit impedance measurements. For the NMR measurements, the strongly anisotropic superconducting properties allowed for orientation in the $ab$ plane \cite{Koutroulakis:2015} to within $\pm0.1^\circ$ by means of a piezoelectric rotator, and orthogonal to the most highly conducting $b$-axis to within $\sim5^\circ$. The sample mass was 0.9 mg, with dimensions 0.8 mm$\times$4.3 mm$\times$0.1 mm. A second crystal gave similar results.

In recounting the NMR measurements, we begin at the intermediate temperatures. The spectrum at $T=78$ K is shown in the lower panel of Fig. \ref{fig:betaDPstructure}. Eight peaks are visible, originating from four inequivalent shifts, and the splitting of each associated with the \C-\C\ dipolar coupling of the spin-labelled ET bridging sites. The spectrum resolves easily the contributions from the two molecules: at the high-frequency end is the 4-peak contribution from one of the inequivalent molecules, and at low frequency is the other. Within a molecule, the frequency shift of the four contributions ($a-d$) are given by
\begin{equation}
\omega_{a-d}=\bar{\omega}\pm\left[\delta\omega^2+d^2\right]^{1/2}\pm d+\omega_{orb},\label{eq:SpectralFit}
 \end{equation}
with $\bar{\omega}\equiv(1/2)(\omega_{out}+\omega_{in})$, $\delta\omega\equiv(\omega_{out}-\omega_{in})$, $d$ the internuclear dipolar coupling, and $\omega_{orb}$ the orbital (chemical) part of the frequency shift. Then the two hyperfine parts of the shift are $K_{out,in}=(\bar\omega)/\omega_0\pm(1/2)\delta\omega/\omega_0$. In contrasting the shifts from the two molecules, we assign the greater to the ``red'', since adjacent to it are the SO$_3$ ligands of the counterions, where the negative charge predominantly resides. Note that details of charge disproportionation in other organic systems are often influenced by coupling to the counterion sublattice \cite{Monceau:2001,Yu:2004,Zorina:2009}. However, near-neighbor repulsive interactions ($V$) would tend to produce a CO even in the case of a higher-symmetry counterion. In the ET compounds, the exhibited CO patterns are stripes; the direction along which they are aligned should depend on parameter details: overlap integrals within and between stacks and strength of the corresponding intrastack (vertical in Fig. \ref{fig:ThetaBetaDP}) and interstack repulsive interactions, as well as coupling to the lattice. The disproportionation seen in the ET salts is commonly associated with a broken symmetry and an insulating state. Here, there is no symmetry breaking and the system is metallic throughout, and both near-neighbor repulsion and counterion interaction are important. Note also that in 2D, there are generally at least two critical $V$ in the extended Hubbard model, with the CO onset occurring at a smaller value for $V$ than the metal-insulator transition. \cite{Seo:2006}.

\begin{figure}[ht]
\includegraphics[width=3.3in]{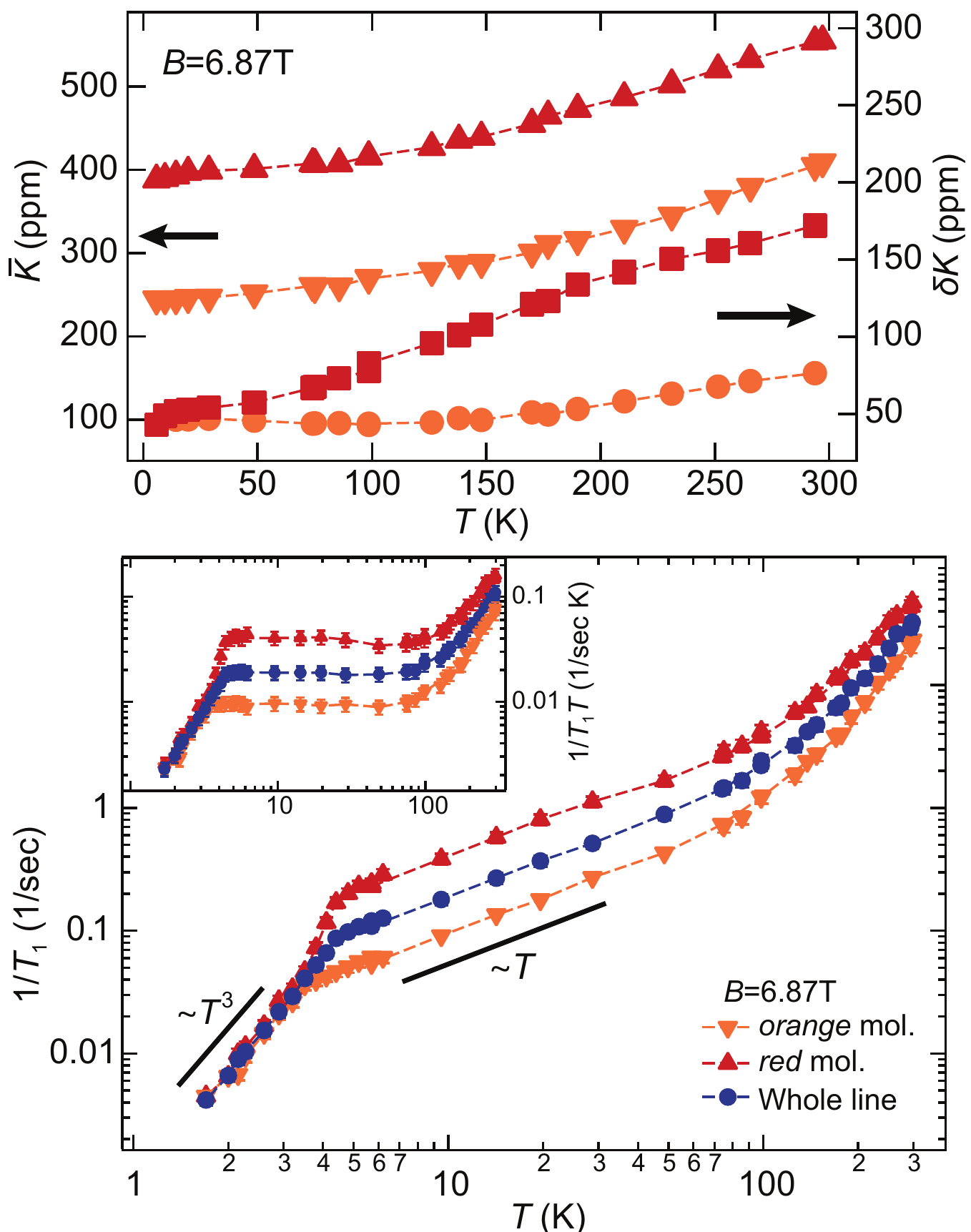} %should be only normal state shift here
\caption{\textit{Top:} Normal-state shifts vs. temperature, averaged for each of the two molecules (left), and difference $\delta K\equiv K_{out}-K_{in}$ vs. temperature (right). The color designation applies to the ``red'' and ``orange'' molecules defined in Fig. \ref{fig:betaDPstructure}. $K_{in}$ is the shift for the \C\ site centered closest to midway between the counterion layers (also Fig. \ref{fig:betaDPstructure}). \textit{Bottom:} In the main panel is \iTone\ vs. $T$, with \iToneT\ vs. $T$ appearing in the inset.} \label{fig:bDPshiftsITone}
\end{figure}

The temperature dependence of the normal-state shifts and relaxation rates for the different sites appears in Fig. \ref{fig:bDPshiftsITone}, where the external field is oriented in the conducting plane and orthogonal to the high-conducting $b$ direction. The data points originating with the two molecules are denoted by the same color scheme as above, red and orange. In the top panel is plotted the average total shift $\overline{K}$ (\textit{left}) and difference $\delta K$ (\textit{right}) for each molecule. Note that the chemical part has not been subtracted out from the total, which is typically of order $K_c\sim$100-150 ppm for ET salts. In the bottom panel appears the spin-lattice relaxation rate \iTone; the colors refer to the average for the two sites on each molecule, and the blue points refer to the average of all sites. \iToneT, which is expected $T$-independent for a Fermi liquid, is plotted in the inset. Both shifts and \iToneT\ increase with higher temperatures. The shifts are anomalous, in the sense that the temperature dependence of the susceptibility is too weak to account for the variation and therefore the hyperfine coupling appears temperature dependent \cite{Dressel:2015}. 

\begin{figure}[ht]
\includegraphics[width=3.3in]{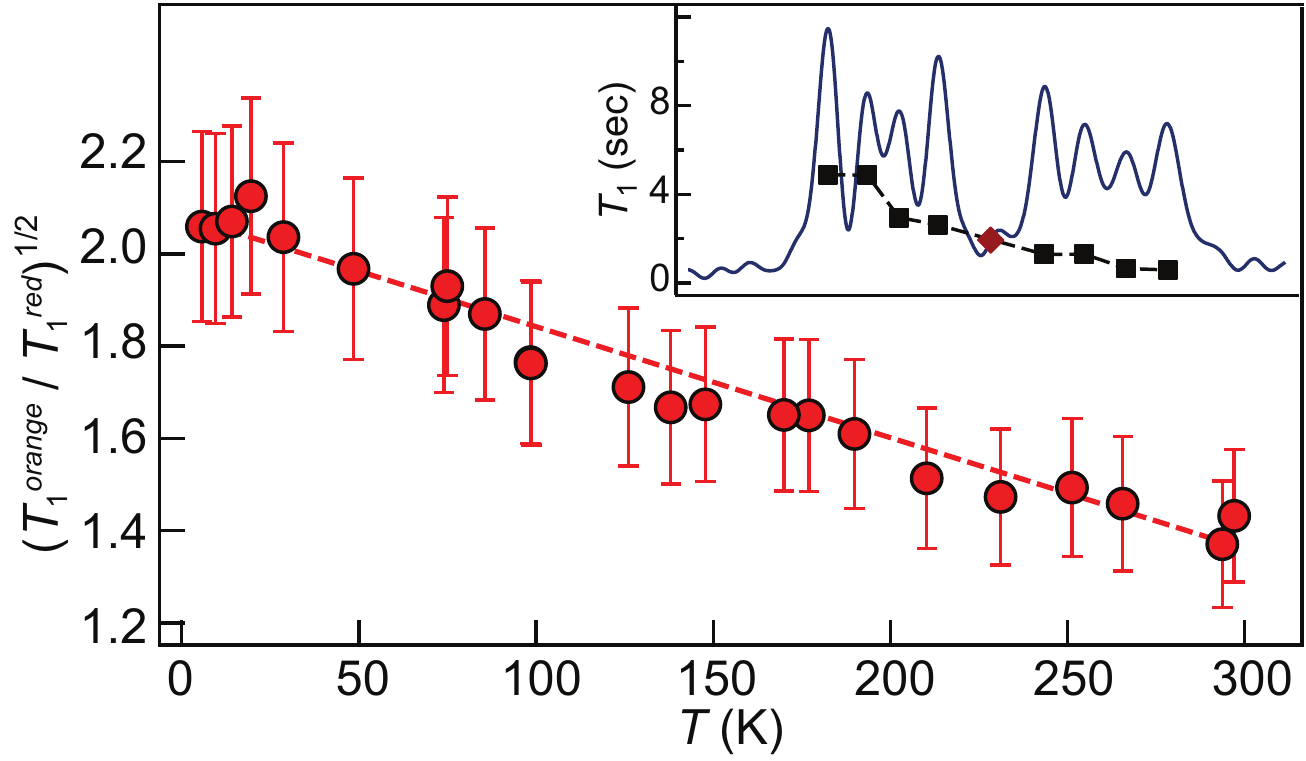}
\caption{Overlaid onto the spectrum in the inset are data (black squares) representing the relaxation time for each of the components. The measurement conditions were $B$=6.87 T, $T$=30 K. The red diamond is the average for the entire spectrum. In the main panel is the temperature variation of $[T_1^{orange}/T_1^{red}]^{1/2}$, which we take to be proportional to the ratio of hole density between red and orange molecules.}\label{fig:HoleRatio}
\end{figure}

Since the hole densities are proportional to the hyperfine fields on the two molecules, the disproportionation can be extracted from the ratio of averaged relaxation rates on the red and orange sites \cite{Chow:2000,Zamborszky:2002},
\begin{equation}
\frac{T_1^{orange}}{T_1^{red}}=\left[\frac{\rho^{red}}{\rho^{orange}}\right]^2,
\end{equation}
where the average hole density is $<\rho>=(1/2)(\rho_{red}+\rho_{orange})$. The result is shown in Fig. \ref{fig:HoleRatio}, which shows a considerable monotonic increase of the disproportionation upon cooling, from $\sim$1.4 at 300 K to $\sim2.0$ at 5 K. The strong and non-trivial $T$ dependence suggests the disproportionation does not result simply from the polar counterions, but rather should be interpreted as a vertical-striped metallic CO state, associated also with intralayer correlations.

\begin{figure}[htb]
\includegraphics[width=3.3in]{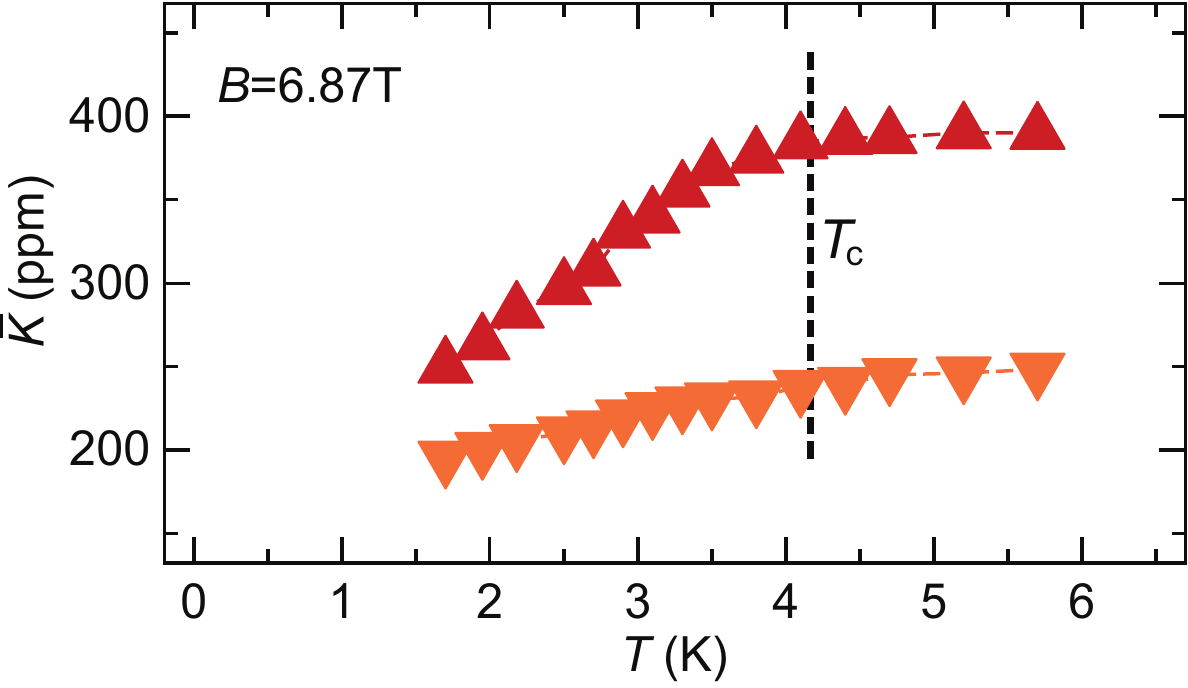}
\caption{$K$ vs. $T$ for $^{13}$C in $\beta''$-(ET)$_2$SF$_5$CH$_2$CF$_2$SO$_3$, measured at low temperature and entering the superconducting state. The average shift for each of the two inequivalent molecules is indicated by the colors, as before. }\label{fig:KsuperBetaDP}
\end{figure}

The very weak temperature dependence of \iToneT\ below 100 K persists down to $T_c$=4.5 K. Below that, the relaxation rate drops abruptly and without exhibiting a Hebel-Slichter peak. The results are compared to a $T^3$ variation in Fig. \ref{fig:bDPshiftsITone}, which is commonly observed in correlated superconductors such as the $\kappa$-phase organic superconductors \cite{DeSoto:1995,Kanoda:1996}, (TMTSF)$_2X$ \cite{Takigawa:1987}, as well as cuprate \cite{Imai:1988} and heavy-fermion superconductors \cite{MacLaughlin:1984,Zheng:2001}. Usually, this is taken as a signature of line nodes and change in sign of order parameter over the Fermi surface, even though there is no specific reason within mean-field theory to find the $T^3$ variation in nodal superconductors except when $T\ll T_c$ \cite{SpecificHeat}. What is also unlike these other superconductors is the absence for enhancement of the normal-state \iToneT\ due to low-energy AF SF, which tend to dominate the relaxation. In the specific example of (TMTSF)$_2$PF$_6$, a greater than five-fold enhancement is observed at temperatures below 20 K \cite{Brown:2008}. We contrast that to the case of the \betadp\ superconductor, in which the enhancement is negligible (Fig. \ref{fig:bDPshiftsITone}).  Further, given that the results here closely resemble other nodal superconductors, such an observation opens the door to whether the physics of pairing in the \betadp\ superconductor would include correlation physics other than, or at least beyond, AF SF. Complementing the relaxation-rate measurements are the low temperature shift data upon entering the SC state, as shown in Fig. \ref{fig:KsuperBetaDP}. The drop in shift is consistent with a decrease in spin susceptibility as for a spin-singlet superconductor. Note that the decrease is much larger for the red molecule than it is for orange, consistent with the interpretation of larger hyperfine fields and spin (carrier) density on the former.

In that context, several questions regarding the pairing in \betaDP\ remain open. To summarize the observations, we have an example of a correlated, metallic CO state which then gives way to superconductivity at low temperature. The SC, at least from the perspective of the NMR results, is consistent with a spin-singlet ground state with nodes on the Fermi surface, but stabilized without evidence for low-energy SF. Prior to this work, there were suggestions that the $\beta''$ materials are an example of SC pairing mediated by CF, first from mean-field calculations on the square lattice and recently as an interpretation of results of IR-conductivity experiments \cite{Merino:2001,Kaiser:2010,Girlando:2014}, and further that a test for the hypothesis would be \iToneT\ unenhanced by SF. The proximity to a nearby CO phase was inferred, in part, from a diversity of ground states observed in other $\beta''$ salts with related counterions. A pertinent example is the salt with counterion SF$_5R$SO$_3$ ($R$=CHFCF$_2$), which undergoes a metal-insulator transition at $\sim$170 K, but which is accompanied by a drop in spin susceptibility \cite{Schlueter:2001}. Given that the system is insulating, the transition is likely accompanied by a lattice distortion to a valence bond solid with singlet ground state. While it is natural to consider that SC is emerging from fluctuations around this state \cite{Dayal:2011}, the \iToneT\ results reported here pose a challenge for this interpretation because there is no drop in \iToneT\ at temperatures approaching $T_c$.

Clearly, direct evidence for charge fluctuations would be beneficial and toward that end two approaches warrant mentioning. First, the available nuclear probes, \C\, $^1$H, $^{19}$F, are all spin $I$=1/2, and thus absent an electric quadrupole moment that couples to the local electric field gradient (EFG). Samples spin-labelled with $^2$H ($I$=1) might correct that deficiency. However, to our knowledge $I>1/2$ nuclei were not previously applied in investigations of CO fluctuations. On the other hand, slow collective charge fluctuations are known to enhance the homogeneous line broadening ($1/T_2$) arising from temporal variations in the paramagnetic shift, in for example $\theta$-(ET)$_2$RbZn(SCN)$_4$, at temperatures greater than the transition to a CO insulator \cite{Chiba:2004}. Similar measurements are currently planned.

\begin{acknowledgments}
The authors acknowledge helpful discussions with Martin Dressel, Steve Kivelson, and Hitoshi Seo. Appreciation is extended to R. Kato and H. Yamamoto for contributing \C\ spin-labelled ET molecules for the crystal growth. The work was supported in part by the National Science Foundation grant number DMR-1410343 (sb). J.W. and H.K. acknowledge the support of the HLD at HZDR, a member of the European Magnetic Field Laboratory (EMFL). JAS acknowledges support from the Independent Research/Development program while serving at the National Science Foundation.
\end{acknowledgments}

\bibliographystyle{apsrev4-1}
%\bibliography{betaDPaoSC}
%merlin.mbs apsrev4-1.bst 2010-07-25 4.21a (PWD, AO, DPC) hacked
%Control: key (0)
%Control: author (72) initials jnrlst
%Control: editor formatted (1) identically to author
%Control: production of article title (-1) disabled
%Control: page (0) single
%Control: year (1) truncated
%Control: production of eprint (0) enabled
%

\end{document}